\documentclass[conference]{IEEEtran}
\IEEEoverridecommandlockouts
\usepackage{threeparttable}
\usepackage{booktabs}

\usepackage{tikz}
\usepackage{pgfplots}
\usepackage{bbding}
\pgfplotsset{width=8cm,compat=1.9}

\usepackage{amssymb}

\usepackage{float}
\usepackage{graphicx}
\usepackage{subcaption}
\usepackage{caption}

\usepackage{multirow}
\usepackage{multicol}
\usepackage{booktabs}
\usepackage{makecell}
\usepackage{tabularx}
\usepackage{longtable}
\usepackage{threeparttable}
\usepackage{wrapfig}
\usepackage{ulem}
\usepackage{colortbl}

\usepackage{csquotes}

\usepackage{soul}

\usepackage{hyperref}
\hypersetup{hidelinks}
\usepackage[ruled,vlined]{algorithm2e}

\usepackage{enumitem}

\usepackage{balance}
\usepackage{pifont}

\usepackage{url}
\usepackage{diagbox}
\usepackage{adjustbox}
\usepackage{breqn}

\usepackage{arydshln}
\usepackage{tcolorbox}
\usepackage[symbol]{footmisc}

\newcommand{\tool}{{UniSRCodec}\xspace}

\def\BibTeX{{\rm B\kern-.05em{\sc i\kern-.025em b}\kern-.08em
    T\kern-.1667em\lower.7ex\hbox{E}\kern-.125emX}}

\begin{document}

\title{\tool: Unified and Low-Bitrate Single Codebook Codec with Sub-Band Reconstruction}

\author{Zhisheng Zhang$^{1*}$\thanks{$^*$Work conducted when the first author was an intern at ModelBest.}, Xiang Li$^1$, Yixuan Zhou$^1$, Jing Peng$^1$, Shengbo Cai$^1$, Guoyang Zeng$^2$, Zhiyong Wu$^{1\dag}$ \\ 
$^1$Shenzhen International Graduate School, Tsinghua University, Shenzhen, China \\
$^2$ModelBest Inc., Beijing, China \\
$^\dag$Corresponding Author}

\maketitle

\begin{abstract}
Neural Audio Codecs (NACs) can reduce transmission overhead by performing compact compression and reconstruction, which also aim to bridge the gap between continuous and discrete signals.
Existing NACs can be divided into two categories: multi-codebook and single-codebook codecs. Multi-codebook codecs face challenges such as structural complexity and difficulty in adapting to downstream tasks, while single-codebook codecs, though structurally simpler, suffer from low-fidelity, ineffective modeling of unified audio, and an inability to support modeling of high-frequency audio. 
We propose the \tool, a single-codebook codec capable of supporting high sampling rate, low-bandwidth, high fidelity, and unified. We analyze the inefficiency of waveform-based compression and introduce the time and frequency compression method using the Mel-spectrogram, and cooperate with a Vocoder to recover the phase information of the original audio. Moreover, we propose a sub-band reconstruction technique to achieve high-quality compression across both low and high frequency bands. Subjective and objective experimental results demonstrate that \tool achieves state-of-the-art (SOTA) performance among cross-domain single-codebook codecs with only a token rate of 40, and its reconstruction quality is comparable to that of certain multi-codebook methods. 
Our demo page is available at \href{https://wxzyd123.github.io/unisrcodec}{https://wxzyd123.github.io/unisrcodec}.

\end{abstract}


\section{Introduction}\label{sec:intro}
The Neural Audio Codec~\cite{encodec, dac} is a compression and recovery technique that converts continuous speech signals into discrete tokens, thereby reducing the cost of audio transmission. In recent years, large audio language models (ALMs)~\cite{qwen2-audio, step-audio2} have garnered considerable attention due to their impressive dialogue capabilities. The speech tokenizer part converts input audio to tokens, feeding them into the LLM. 
Moreover, the information entropy of tokens per second is larger, the better it is for LLM to extract information.

Existing NACs can be broadly categorized into multi-codebook and single-codebook codecs based on the number of utilized codebooks. Multi-codebook codecs have dominated prior research. By hierarchically leveraging multiple codebooks, \textit{e.g.}, Residual Vector Quantization (RVQ), they achieve high-fidelity audio reconstruction. However, such codecs produce multi-level token sequences, which introduce complexity for adapting into ALMs or text-to-speech systems. In recent years, single-codebook codecs have been explored due to their architectural simplicity, \textit{e.g.}, BigCodec~\cite{bigcodec}, WavTokenizer~\cite{wavtokenizer}, and UniCodec~\cite{unicodec}. However, previous single-codebook codecs suffer from two main limitations: (1) \textbf{Universality.} They perform inferior modeling capabilities for general audio, \textit{e.g.}, BigCodec. (2) \textbf{High-Frequency Modeling.} They often operate at low sampling rates,  with high bandwidth or computational resources heavy. Lower sampling rates, such as 24kHz, may suffice for speech content but fail short for music or general audio with lower perceptual quality than high sampling rates. Moreover, higher sampling rates simultaneously pose substantial challenges for unified audio representation and modeling.

To address above limitations, we propose a neural audio codec named \textbf{Uni}fied \textbf{S}ub-band \textbf{R}econstruction \textbf{Codec} (\tool), a {\it \uline{high-sampling-rate}}, {\it \uline{low-bitrate}}, {\it \uline{high-fidelity}}, and {\it \uline{unified}} {\it \uline{single-codebook}} audio codec with {\it \uline{training-lightweight}} requirements. The single-codebook design is intended to better align with downstream tasks. High-frequency modeling enables richer, more natural audio quality and perceptual fidelity, and allows the codec to effectively model general audio types. The low-bitrate requirement demands that each token carry as much information as possible, reflecting the information density. High-fidelity ensures minimal information loss during the discretization process, which is essential for compression. Unified capability requires the codec's ability to model cross-domain audio. Moreover, training-lightweight represents that the codec is resource-friendly for training.

Waveform-based techniques~\cite{wavtokenizer, unicodec} that compress the time domain often retain redundant information and exhibit worse modeling of the spectral domain. To mitigate this, we aim to propose a compression strategy with better information density per token using Mel-spectrograms as the reconstruction representation. During compression, we intentionally omit phase information, thereby allocating bandwidth more efficiently to perceptually critical components. The discarded phase is later recovered during audio reconstruction via a neural vocoder, which synthesizes high-quality waveforms from the Mel representations. Using this approach, we achieve high-fidelity audio compression and reconstruction at a token rate of only 40 and an ultra-low bitrate of 0.52 kbps. Additionally, compared to UniCodec~\cite{unicodec}, the training process is computationally lightweight, requiring only 8 NVIDIA RTX 4090 GPUs for approximately 12 hours. Experimental results demonstrate that our method achieves state-of-the-art (SOTA) performance among single-codebook codecs in cross-domain audio modeling, while achieving reconstruction fidelity comparable to that of multi-codebook approaches.

\begin{figure*}[t]
    \centerline{
    \includegraphics[width=0.85\textwidth]{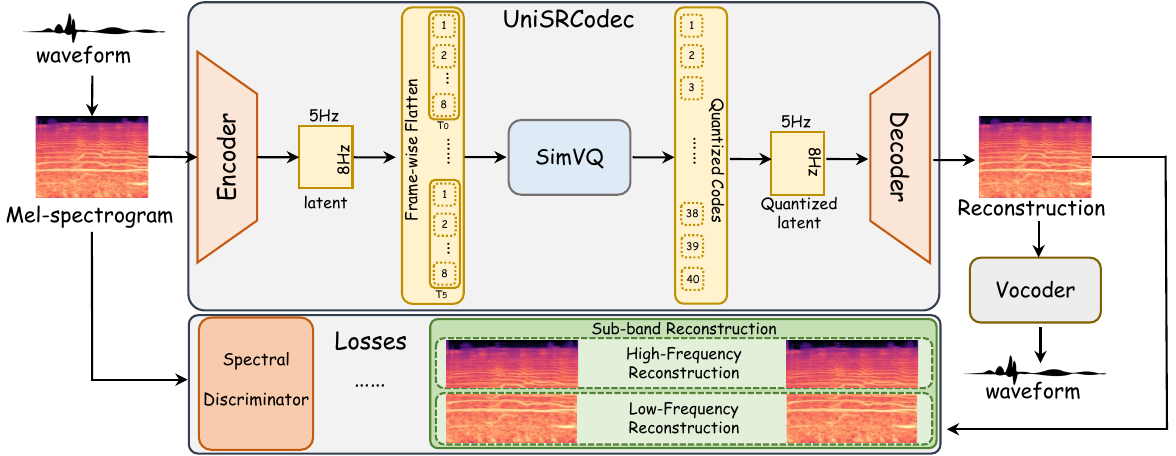}}
    \caption{The architecture and training procedure of the \tool.}
    \label{fig_workflow}
\end{figure*}

Our main contributions are summarized as follows:
\begin{itemize}
    \item We introduce \tool, a high-sampling-rate, ultra-low-bitrate, and unified single-codebook audio codec that achieves high-quality audio modeling by exploiting both time and frequency domain representations.
    \item We propose a sub-band reconstruction approach to better model both low and high frequency information.
    \item We evaluate \tool on diverse cross-domain datasets, including speech, music, and general sound, and demonstrate the SOTA performance among single-codebook codecs using only 40 tokens per second.
\end{itemize}

\section{Related Work}
\subsection{Single-Codebook Codec}
Single-codebook codecs are structurally simple and can be easily adapted to downstream tasks or employed as speech tokenizers for input into LLMs. BigCodec~\cite{bigcodec} is a low-bitrate speech codec that enhances compression capability by scaling up model parameters. WavTokenizer~\cite{wavtokenizer} pioneers the integration of multi-level codebooks into a single codebook. By initializing the codebook with K-means clustering and incorporating attention mechanisms in the decoder, WavTokenizer enables high-fidelity speech reconstruction at low bitrates. Building upon WavTokenizer, UniCodec~\cite{unicodec} proposes the use of a Mixture-of-Experts (MoE) architecture combined with a large domain-adaptive codebook of size 16384 to model unified audio. However, these single-codebook codecs suffer from the domain constraints, \textit{e.g.}, BigCodec, or limited sampling rates, \textit{i.e.}, 16kHz or 24kHz, which constrain their ability to faithfully reconstruct high-fidelity audio. Moreover, their reconstruction quality under ultra-low-bandwidth conditions can still be improved on unified audio.

\subsection{Spectral-based NAC}
Some prior study has explored compression in the frequency domain. APCodec~\cite{apcodec} proposes a frequency-domain compression approach that applies 1D convolutions separately to the magnitude and phase spectra obtained from the Short-Time Fourier Transform (STFT) of the waveform. However, this method compresses only in the frequency domain and lacks explicit modeling of temporal information. FunCodec~\cite{funcodec} extends this by employing 2D convolutions to jointly compress both time and frequency dimensions. While its multi-codebook structure limits its practical applicability. MelCap~\cite{melcap} is a work closely related to ours and resembles FunCodec, but replaces the RVQ with a single codebook. Notably, MelCap leverages pre-trained VGG weights from the image domain to construct Mel-spectrograms, {\it \uline{lacking specialized design considerations in the audio domain}}. Moreover, it operates at a relatively high-bandwidth with a token rate of 260 and a bitrate of 3.4 kbps. Therefore, we aim to develop a high-fidelity single-codebook codec tailored for the audio domain that operates effectively under even lower bandwidth conditions.

\section{\tool Design}
In this section, we explain why we select the Mel-spectrogram as the input representation, followed by a detailed description of the \tool architecture.

\subsection{Why Regarding Mel-spectrograms as Input?}
In this section, we illustrate reasons why we utilize mel-spectrograms as input features.

\textbf{Information Density.} Consider an audio segment of length 65536 sampled at 44.1kHz. After applying STFT with a hop size of 512, the Mel-spectrogram has dimensions of $128 \times 128$. When flattened, this results in a sequence of length 16384, approximately 25\% of the original waveform length. Although the Mel-spectrogram discards phase information, advanced vocoders have demonstrated the ability to accurately reconstruct phase from Mel-spectrograms alone~\cite{bigvgan}. Therefore, using Mel-spectrograms as input enables a more efficient and information-rich representation compared to raw waveforms.

\textbf{Quadratic Compression Efficiency.} Unlike the raw waveform, which is one-dimensional, the Mel-spectrogram is inherently two-dimensional. Compressing the waveform by $n$ times corresponds to reducing resolution only along the temporal dimension. In contrast, compressing the Mel-spectrogram by a factor of $n$ simultaneously reduces resolution in both time and frequency, yielding an overall compression ratio of $n \times n = n^2$. This quadratic gain in compression efficiency is a key enabler for achieving an ultra-low token rate of 40.

\textbf{Time and Frequency Domain Compression.} Compression applies solely to the temporal information, which may lead to loss of spectral energy details. To achieve high-fidelity audio reconstruction, it is essential to preserve information across both time and frequency domains.

\subsection{Architecture}
\tool extracts the Mel-spectrogram from the input audio and reconstructs it using an encoder–quantizer–decoder architecture. The reconstructed Mel-spectrogram is then passed to a pre-trained BigVGAN-v2~\cite{bigvgan} to recover the audible waveform. The workflow is shown in Figure \ref{fig_workflow}.

\textbf{Encoder.} Our encoder and decoder architectures are adapted from Open-MagViT2~\cite{open-magvit2}. The encoder employs a fully convolutional 2D architecture composed of multiple residual blocks (ResBlocks). Each ResBlock consists of two GroupNorm layers with activation functions and two convolutional layers. The input channel is a single one for the mel-spectrogram. Throughout the encoder, the number of channels is progressively increased to $[128, 256, 512]$, while spatial resolution is reduced via strided convolutions. Specifically, temporal dimensions are downsampled by factors of $[2, 2, 4]$ and frequency dimensions by $[2, 2, 4]$, resulting in an overall compression ratio of $16 \times 16$. During training, a $128 \times 128$ Mel-spectrogram is compressed into an $8 \times 8$ latent representation.

\textbf{Quantizer.} Since the latent representation from the encoder is two-dimensional, it must be flattened into a one-dimensional vector to be compatible with quantization. We consider two flattening strategies: {\it \uline{band-wise}} and {\it \uline{frame-wise}}. Band-wise flattening preserves all temporal information by concatenating time sequences for each Mel frequency bin, whereas frame-wise flattening concatenates all frequency bins within each time frame. In \tool, we adopt the frame-wise flattening strategy to preserve coherent frequency dynamics across time, ensuring that each frame evolves as a complete spectral unit. The performance of these two flattening strategies is illustrated in Section \ref{section:exp_ablation}.

\textbf{Decoder.} The decoder mirrors the encoder architecture symmetrically. Starting from the quantized codebook vectors, it reconstructs the Mel-spectrogram. The decoder begins with 512 input channels and progressively reduces the channel count to $[256, 128, 1]$. Concurrently, it upsamples the spatial dimensions: the temporal axis is expanded by factors of $[4, 2, 2]$ and the frequency axis by $[4, 2, 2]$, ultimately reconstructing the latent representation back to the original $128 \times 128$ Mel-spectrogram size.

\section{Training Procedure}
In this section, we propose the sub-band reconstruction strategies training process and objective functions of \tool. Training \tool primarily involves the encoder, quantizer, and decoder. 

\subsection{Sub-band Reconstruction}
Since the modeling of the Mel spectrogram involves information across different frequency bands, we first reconstruct the entire Mel spectrogram and observe that although high-frequency signals are well modeled, the low-frequency signals degrade to some extent. Moreover, the information in low-frequency regions of the Mel spectrogram is more fine-grained and therefore deserves greater attention and learning from the model. Based on this observation, we propose a sub-band reconstruction approach. Assuming the Mel spectrogram has $m$ Mel bins and the number of time frames after the STFT transformation is $t$, the input data is represented as a vector $x \in \mathbb{R}^{m \times t}$. We divide the frequency axis into two halves: the first half corresponds to the low-frequency signal $x_{\text{low}} \in \mathbb{R}^{\frac{m}{2} \times t}$, and the latter half corresponds to the high-frequency signal $x_{\text{high}} \in \mathbb{R}^{\frac{m}{2} \times t}$. We compute the L1 loss separately for the reconstructed low-frequency signal $\hat{x}_{\text{low}}$ and high-frequency signal $\hat{x}_{\text{high}}$, and use their weighted average as the overall reconstruction loss as Eq. (\ref{eq:recon}). We will validate this approach in Section \ref{section:exp_ablation}.
\begin{equation}
    \mathcal{L}_{\text{sr}} = \frac{\left( \alpha_{\text{low}} |x_{\text{low}} - \hat{x}_{\text{low}}|_1 + \alpha_{\text{high}} |x_{\text{high}} - \hat{x}_{\text{high}}|_1 \right)}{\alpha_{\text{low}} + \alpha_{\text{high}}}. \label{eq:recon}
\end{equation}

\subsection{Mel-Spectrogram Reconstruction Training}
During this process, we train the encoder, quantizer, and decoder, intending to generate high-quality Mel-spectrograms. Moreover, the training of Mel-spectrograms also benefits from the inclusion of a discriminator. Otherwise, training the model alone would lead to over-smoothing artifacts in Section \ref{section:exp_ablation}.

\textbf{Reconstruction Loss.} We utilize our proposed sub-band reconstruction $\mathcal{L}_{\text{sr}}$ with $\alpha_{\text{low}} = 2$ and $\alpha_{\text{high}} = 1$.

\textbf{Discriminator Loss.} Introducing a discriminator helps the generator learn fine-grained spectral details. We adopt the multi-band multi-scale STFT discriminator architecture from DAC~\cite{dac} but remove the multi-band and multi-scale components, as the Mel-spectrogram input already has fixed frequency resolution and scale. This loss is denoted as $\mathcal{L}_{\text{disc}}$.

\textbf{Adversarial Loss} Following DAC, we use an adversarial loss $\mathcal{L}_{\text{adv}}$ using the spectral discriminator, along with a feature matching loss in the frequency domain, denoted as $\mathcal{L}_{\text{fm}}$.

\textbf{Codebook Loss.} We employ SimVQ~\cite{simvq} as the single-codebook quantizer and use a commitment loss $\mathcal{L}_{\text{cm}}$ to optimize the codebook vectors achieving higher utilization.

\textbf{Training Objective.} We optimize the aforementioned loss terms, resulting in the overall training objective in Eq. (\ref{eq_s1}).
\begin{equation}
\mathcal{L} = \lambda_{\text{sr}} \mathcal{L}_{\text{sr}} + \lambda_{\text{disc}} \mathcal{L}_{\text{disc}} + \lambda_{\text{adv}} \mathcal{L}_{\text{adv}} + \lambda_{\text{fm}} \mathcal{L}_{\text{fm}} + \lambda_{\text{cm}} \mathcal{L}_{\text{cm}}, \label{eq_s1}
\end{equation}
where the coefficients $\lambda_{\text{sr}}, \lambda_{\text{disc}}, \lambda_{\text{adv}}, \lambda_{\text{fm}}, \lambda_{\text{cm}}$ are set to 15, 1, 1, 1, 1, respectively. The selection of these hyperparameters closely follows established practices in prior work~\cite{llasa, dac}, with only minor modifications to achieve optimal performance.

\begin{table*}[t]
    \caption{Objective evaluation of the reconstruction performance on general, music, and speech domain test datasets. ``Mel-44'' and ``Mel-16'' represent the Mel Distance on both high and low frequency as same as ``STFT-44'' and ``STFT-16'' on STFT Distance. \textbf{Bold} denotes the best performance in the single-codebook NACs and \uline{underline} reflects the second-best performance.}
    \resizebox{0.98\linewidth}{!}{
    \scalebox{1.1}{
    \begin{threeparttable}
    \begin{tabular}{ccccccccccccccccccccc}
    \toprule[1pt]\midrule[0.3pt]
    \multirow{2}{*}{\textbf{Models}}
    & \multicolumn{4}{c}{\textbf{Attribution}}
    & \multicolumn{4}{c}{\textbf{AudioSet-Eval}}
    & \multicolumn{4}{c}{\textbf{MusicDB test}}
    & \multicolumn{3}{c}{\textbf{LibriTTS test}}
    \\
    \cmidrule(r){2-5} \cmidrule(lr){6-9} \cmidrule(lr){10-13} \cmidrule(lr){14-17}
    & Unified & TPS & kbps/Nq & SR
    & Mel-44($\downarrow$) & STFT-44($\downarrow$)
    & Mel-16($\downarrow$) & STFT-16($\downarrow$)
    & Mel-44($\downarrow$) & STFT-44($\downarrow$)
    & Mel-16($\downarrow$) & STFT-16($\downarrow$)
    & STOI($\uparrow$) & PESQ($\uparrow$)
    \\
    \midrule
    \rowcolor{gray!10}\multicolumn{15}{c}{\textit{Vocoder Reconstructs with Ground Truth Mel-spectrograms}} \\
    \midrule
    BigVGAN~\cite{bigvgan}
        & \ding{52} & - & - & 44.1
        & 0.417 & 1.713 & 0.363 & 1.683
        & 0.380 & 1.334 & 0.350 & 1.263
        & 0.993 & 4.186 \\
    \midrule
    \rowcolor{gray!10} \multicolumn{15}{c}{$>$ \textit{100 token rate}} \\
    \midrule
    DAC~\cite{dac}
        & \ding{52} & 900 & 9/9q & 44.1 
        & \textbf{0.654} & \textbf{1.958} & \textbf{0.625} & \textbf{1.842}
        & \textbf{0.651} & 1.634 & 0.665 & \textbf{1.462}
        & \textbf{0.972} & \textbf{3.900} \\
    Encodec~\cite{encodec}
        & \ding{52} & 600 & 6/8q & 24
        & 1.315 & 5.030 & 0.889 & 2.271
        & 1.372 & 4.705 & 1.086 & 2.020
        & \uline{0.943} & \uline{2.819}\\
    Encodec~\cite{encodec}
        & \ding{52} & 300 & 3/4q & 24
        & 1.413 & 5.134 & 1.017 & 2.448
        & 1.463 & 4.769 & 1.203 & 2.134
        & 0.904 & 2.116 \\
    SNAC~\cite{snac}
        & \ding{52} & 240 & 2.88/4q & 44.1
        & 0.828 & 2.145 & 0.863 & 2.234
        & 0.788 & 1.673 & 0.845 & 1.645
        & 0.925 & 2.561  \\
    MelCap~\cite{melcap}
        & \ding{52} & 260 & 3.4/1q & 44.1
        & 0.817 & 2.229 & 0.873 & 2.409
        & 0.796 & 1.813 & 0.896 & 1.870
        & 0.888 & 1.802 \\
    \hdashline
    \textbf{\tool-L}
        & \ding{52} & 176 & 2.29/1q & 44.1
        & \uline{0.729} & \uline{2.049} & \uline{0.692} & \uline{2.093}
        & \uline{0.656} & \textbf{1.543} & \textbf{0.638} & \uline{1.529}
        & 0.941 & 2.727 \\
    \midrule
    \rowcolor{gray!10} \multicolumn{15}{c}{$\le$ \textit{100 token rate}} \\
    \midrule
    DAC~\cite{dac}
        & \ding{52} & 100 & 1/1q & 44.1
        & \uline{1.187} & \uline{2.588} & 1.282 & 2.752
        & \uline{1.276} & \uline{2.195} & 1.474 & 2.270
        & 0.763 & 1.308 \\
    BigCodec~\cite{bigcodec}
        & \ding{55} & 80 & 1.04/1q & 16
        & 2.250 & 7.336 & 1.366 & 3.024
        & 1.958 & 6.639 & 1.031 & 2.003
        & \textbf{0.943} & \uline{2.700} \\
    TAAE~\cite{taae}
        & \ding{55} & 50 & 0.7/1q & 16
        & 2.999 & 7.896 & 2.385 & 4.314
        & 2.490 & 7.006 & 1.746 & 2.927
        & 0.890 & 1.787 \\
    WT-Speech~\cite{wavtokenizer}
        & \ding{55} & 75 & 0.9/1q & 24
        & 1.393 & 5.179 & 1.026 & 2.572
        & 1.341 & 4.700 & 0.997 & 1.923
        & 0.922 & 2.566 \\
    WT-MA~\cite{wavtokenizer}
        & \ding{55} & 75 & 0.9/1q &24
        & 1.396 & 5.132 & 0.985 & 2.453
        & 1.390 & 4.689 & 1.044 & 1.977
        & 0.857 & 1.747 \\
    WT-Unified~\cite{wavtokenizer}
        & \ding{52} & 40 & 0.48/1q & 24
        & 1.505 & 5.242 & 1.130 & 2.634
        & 1.558 & 4.770 & 1.255 & 2.110
        & 0.875 & 1.912 \\
    UniCodec~\cite{unicodec}
        & \ding{52} & 75 & 1.3/1q & 24
        & 1.376 & 5.169 & \uline{0.903} & \uline{2.401}
        & 1.352 & 4.713 & \uline{0.943} & \uline{1.858}
        & \uline{0.940} & \textbf{2.870} \\
    \hdashline
    \textbf{\tool-B}
        & \ding{52} & 40 & 0.52/1q & 44.1
        & \textbf{0.904} & \textbf{2.250} & \textbf{0.900} & \textbf{2.330}
        & \textbf{0.882} & \textbf{1.747} & \textbf{0.893} & \textbf{1.768}
        & 0.875 & 1.836 \\
    \midrule[0.3pt]\bottomrule[1pt]
    \end{tabular}
    \begin{tablenotes}
        \item (1){ \textbf{WT-Speech}: WavTokenizer~\cite{wavtokenizer} on the speech domain.
            (2) \textbf{WT-MA}: WavTokenizer on the music and audio domain.
            (3) \textbf{WT-Unified}: WavTokenizer is unified.
            (4) \textbf{Nq}: the number of quantizer(s).}
    \end{tablenotes}
    \end{threeparttable}
    }
    }
    \label{table:exp_main}
\end{table*}

\section{Experiments and Analyses}

\subsection{Experimental Setup}

\textbf{Datasets.} The training set covers nearly 10000 hours of cross-domain data. For the speech domain, we employ the VCTK~\cite{vctk}, LibriTTS~\cite{libritts}, and Common Voice~\cite{common_voice}. For the music type, we use the MusicDB~\cite{musdb18} and Jamendo~\cite{jamendo}. For the general audio, we use the AudioSet~\cite{audioset}. For the test set, we utilize the LibriTTS test-clean, MUSDB test, and AudioSet eval, each with 1000 samples per domain~\cite{dac, unicodec}.

\textbf{Metrics.} 
For cross-domain data, we employ optimal metrics. For music and general sound, following DAC~\cite{dac}, we compute the Mel-spectrogram distance and STFT distance by calculating the L1 loss between the mel-spectrograms and linear-spectra of the original and reconstructed audio in the high (44kHz) and low (16kHz) frequency components, respectively. For speech evaluation, following UniCodec~\cite{unicodec}, we select speech-related metrics, including STOI and PESQ, to assess the generation quality of the reconstructed speech.

For quantization metrics, we utilize the Tokens Per Second (TPS)~\cite{unicodec} and bandwidth (kbps). TPS denotes the number of tokens for modeling one second of audio. Bandwidth, measured in kilobits per second (kbps), represents the data rate required to transmit the quantized audio tokens and reflects the codec's efficiency in terms of transmission or storage cost. For subjective evaluation, we perform a MUSHRA-inspired listening test~\cite{dac} in Section \ref{section:subjective}.

\textbf{Baselines.} We consider SOTA NACs, including DAC~\cite{dac}, Encodec~\cite{encodec}, SNAC~\cite{snac} with multi-codebook and BigCodec~\cite{bigcodec}, MelCap~\cite{melcap}, TAAE~\cite{taae}, WavTokenizer~\cite{wavtokenizer}, UniCodec~\cite{unicodec} with single-codebook.

\textbf{Training Details.} We train the \tool on 8 NVIDIA 4090 for $100000 (\times 8)$ steps using the AdamW optimizer with the initial learning rate as $1 \times 10^{-4}$ and batch size 20. 

\subsection{High-Frequency Data Training}
To achieve high-frequency modeling, we require substantial high-resolution data. Upsampling low-sampling-rate audio can result in the loss of high-frequency components. Therefore, we perform preliminary filtering on the training data. First, we compute the mean energy for each Mel band, then search downward from the highest-frequency band. If a band's mean energy exceeds a predefined threshold, we consider all lower-frequency bands beneath it to contain valid energy information, and the sampling rate corresponding to this band is regarded as the audio's native sampling rate. We empirically set the energy threshold to -60dB and, following DAC~\cite{dac}, select all audio data whose true bandwidth exceeds the Nyquist frequency (22.05kHz) as training data. This helps the model's ability to model high-frequency signals~\cite{dac}.

\subsection{Evaluation on Cross-Domain Datasets}
In this section, we evaluate \tool's performance on speech, music, and general sound datasets. 

Table \ref{table:exp_main} presents the evaluation results of our method and the baselines on the cross-domain dataset. We provide two variants of \tool: ``\tool-B'' denotes the base version with an ultra-low token rate of 40, and ``\tool-L'' refers to a slightly larger-bitrate variant designed to align with multi-codebook NACs. We observe that \tool-B achieves SOTA performance among single-codebook methods on both music and general audio domains, faithfully reconstructing both high and low frequency signals. Compared to UniCodec~\cite{unicodec}, our approach operates at a lower bitrate and bandwidth while still enabling high-fidelity reconstruction of high-frequency components. According to the metrics ``Mel-44'' and ``STFT-44'', our proposed \tool can model high-frequency signals better, indicating better performance when modeling high-fidelity music and general audio. Furthermore, \tool-L outperforms multi-codebook methods, \textit{e.g.}, SNAC and Encodec, at an even lower bitrate, and matches or even surpasses DAC in performance, exceeding DAC on two metrics in the music domain, \textit{i.e.}, ``STFT-44'' and ``Mel-16''.

In the speech domain, \tool-B shows somewhat lower modeling capability compared to UniCodec. This is primarily because UniCodec’s training data consists overwhelmingly of $\sim$700000-hour speech, and UniCodec incorporates semantic learning, which enhances semantic fidelity at the cost of increased training complexity. From the perspective of \tool-L, when the model’s sampling rate is set to 24kHz, yielding a token rate of ~90, it achieves speech modeling performance comparable to UniCodec while significantly outperforming it in general audio modeling.

This experiment not only demonstrates \tool's superiority in modeling cross-domain data at a low token rate of 40, especially on music and general sound, but also validates its scalability, as it surpasses the multi-codebook codec SNAC with a token rate of 176.

\begin{table}[t]
    \caption{The ablation study on AudioSet. ``w/o'' represents training without the component while ``w'' denotes with it.}
    \centering
    \resizebox{0.9\linewidth}{!}{
    \scalebox{1.1}{
    \begin{tabular}{ccccccccccccccc}
    \toprule[1pt]\midrule[0.3pt]
    \multirow{2}{*}{\textbf{Method}}
    & \multicolumn{4}{c}{\textbf{AudioSet}}
    \\
    \cmidrule(r){2-6} 
    & Mel-44($\downarrow$) & STFT-44($\downarrow$)
    & Mel-16($\downarrow$) & STFT-16($\downarrow$) \\
    \midrule
    \textbf{\tool}
        & \textbf{0.904} & \textbf{2.250} & \textbf{0.900} & \textbf{2.330} \\
    \hdashline
    w/o Discriminator
        & 1.261 & 2.493 & 1.278 & 2.645 \\
    w/o Sub-band
        & 0.909 & 2.247 & 0.922 & 2.347 \\
    w Scheduler 
        & 0.929 & 2.287 & 0.927 & 2.364 \\
    w/o Frame-wise Flatten
        & 0.930 & 2.275 & 0.932 & 2.363 \\
    \midrule[0.3pt]\bottomrule[1pt]
    \end{tabular}
    }
    }
    \label{table:exp_ablation}
\end{table}

\subsection{Ablation Study}\label{section:exp_ablation}
In this section, we explore the ability of each component.

\textbf{Discriminator.} In the \tool, we design a lightweight discriminator to enhance the codec’s learning of fine-grained mel-spectrogram details. This is achieved by performing temporal and spectral downsampling on the input and computing feature matching loss based on the features extracted at each downsampling stage. As shown in Table \ref{table:exp_ablation}, removing the discriminator and its associated loss functions, retaining only the reconstruction loss and codebook learning loss, leads to a significant performance degradation, with synthesized audio exhibiting audible electronic artifacts.  

\textbf{Sub-band Reconstruction.} Sub-band reconstruction aims to amplify the weighting of low-frequency signal components, thereby improving the model’s capacity to learn low-frequency features. When we replace our proposed strategy with a conventional reconstruction loss, \textit{i.e.}, computing L1 loss over the entire mel-spectrogram, the reconstruction quality in the low-frequency range deteriorated. Specifically, the ``Mel-16'' metric increases from the original 0.901 to 0.922, validating the effectiveness of our method for low-frequency modeling. This strategy is specifically tailored to the unique characteristics of our time-frequency compression approach, which enables independent processing of distinct frequency bands.  

\textbf{Scheduler.} During training, the learning rate is fixed at $1 \times 10^{-4}$ due to the model’s relatively rapid feature learning capability. We test with adding an exponential scheduler with the decay rate as $0.999976974$, reducing the learning rate from $1 \times 10^{-4}$ to $1 \times 10^{-5}$ after 100000 steps. Table \ref{table:exp_ablation} demonstrates that incorporating this scheduler resulted in a performance decline, \textit{e.g.}, from 0.903 to 0.929 in the ``Mel-44'' metric, indicating that a constant learning rate may better suit \tool to achieve optimal learning performance.

\textbf{Frame-wise and Band-wise Flattening.} Since our input is the mel-spectrogram, the encoder output remains a 2D vector. To facilitate codebook lookup, the vector needs to be flattened into a 1D vector before being fed into the codebook. The flattening strategy can follow two approaches: (1) {\it \uline{Frame-wise Flattening}}, which concatenates frequency information within each time frame, or (2) {\it \uline{Band-wise Flattening}}, which concatenates temporal points across frequency bands. We test the band-wise flattening strategy, and the results, as shown in Table \ref{table:exp_ablation}, demonstrate a performance degradation across all metrics on the AudioSet dataset. This is because flattening across time steps during training could lead to inconsistencies during inference, as variable audio lengths might degrade model performance. In contrast, frame-wise flattening ensures consistency between training and inference, since the mel-bin dimension, \textit{i.e.}, frequency axis remains fixed. Based on this analysis, we adopt frame-wise flattening in \tool.


\subsection{Subjective Evaluation}\label{section:subjective}
\begin{figure}[t]
    \centerline{
    \includegraphics[width=0.35\textwidth]{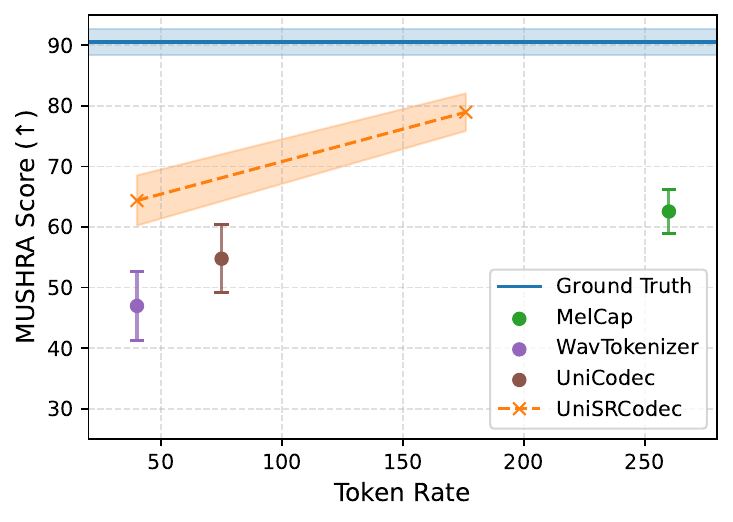}}
    \caption{Subjective evaluation of MUSHRA scores with 95\% confidence intervals vs token rate.}
    \label{fig:subjective}
\end{figure}

To evaluate the perceptual performance of \tool, we conduct a MUSHRA listening test. We randomly select three audio samples from each of three domains, totaling nine audio clips, which are subjectively scored by ten experts based on reconstruction quality. Figure \ref{fig:subjective} presents the average results across the three domains, where \tool-B and \tool-L achieve the second-highest and highest subjective reconstruction scores, respectively.

In audio and music domains, \tool surpasses UniCodec~\cite{unicodec}, the SOTA unified single-codebook model. Specifically, in the audio domain, \tool-B and \tool-L score 53.433 and 72.667, respectively, outperforming UniCodec's 46.900. In the music domain, the improvement is more pronounced: \tool-B and \tool-L achieve average MUSHRA scores of 62.833 and 80.967, respectively, outperforming both MelCap (59.900) and UniCodec (33.933). For speech, \tool-B and \tool-L achieve average MUSHRA scores of 76.867 and 83.233, respectively, demonstrating performance comparable to UniCodec's 83.467. Additionally, we find that the base version with a 40-token rate already outperforms MelCap's 260-token rate, achieving SOTA performance among single-codebook models, which is attributed to our design and the proposed loss function. Moreover, the better performance of \tool than UniCodec comes from the capability of high-frequency modeling.

\begin{table}[t]
    \caption{The cross-domain downstream classification tasks.}
    \centering
    \resizebox{0.9\linewidth}{!}{
    \scalebox{1.1}{
    \begin{tabular}{ccccccccccccccc}
    \toprule[1pt]\midrule[0.3pt]
    \multirow{2}{*}{\textbf{Model}}
    & \multirow{2}{*}{\textbf{TPS}}
    & \multicolumn{2}{c}{\textbf{Sound}}
    & \textbf{Music} & \textbf{Speech}
    \\
    \cmidrule(r){3-4} \cmidrule(lr){5-5} \cmidrule(lr){6-6} 
    & & UrbanSound-8k & ESC-50
    & GTZAN & CREMA-D \\
    \midrule
    \rowcolor{gray!10} \multicolumn{6}{c}{\textit{Continuous Representation}} \\
    \midrule
    WavLM~\cite{wavlm}
        & -& 0.53 & 0.32 & 0.48 & 0.45 \\
    \midrule
    \rowcolor{gray!10} \multicolumn{6}{c}{\textit{Discrete Representation}} \\
    \midrule
    WavTokenizer~\cite{wavtokenizer}
        & 40 & 0.33 & 0.17 & 0.40 & 0.39 \\
    \tool
        & 40 & \textbf{0.40} & \textbf{0.19} & \textbf{0.40} & \textbf{0.42} \\
    \midrule[0.3pt]\bottomrule[1pt]
    \end{tabular}
    }
    }
    \label{table:exp_downstream}
\end{table}

\subsection{Downstream Understanding Tasks}

In this section, we evaluate the performance of the \tool in downstream understanding tasks. The encoder and quantizer of the codec typically serve as the discretization strategy for ALMs. Therefore, the ability of NACs to understand audio, particularly general audio, also deserves attention. We adopt the xares benchmark~\cite{xares}, feeding the embeddings obtained after the quantizer of the codec into an MLP provided by xares to adapt to various downstream tasks. Performance scores across different tasks are standardized, with higher scores indicating better performance. We select four distinct and cross-domain tasks: UrbanSound-8k for urban environmental sound classification, ESC-50 for various environmental sound classification, GTZAN Genre for music genre classification, and CREMA-D for emotion recognition. For continuous representations, we select WavLM~\cite{wavlm}, which is pre-trained on large-scale data. For discrete representations, we choose WavTokenizer, which has the same token rate as \tool. For \tool, we flatten the embeddings by frame-wise strategies to match the input of the MLP layer.

Table \ref{table:exp_downstream} presents downstream performances. The proposed \tool outperforms WavTokenizer on all four tasks and achieves performance comparable to the continuous representation model WavLM on certain CREMA-D. This experiment also validates the effectiveness of the \tool for general audio understanding in downstream tasks.

\section{Conclusion}
In this paper, we propose a unified and low-bitrate single-codebook \tool. For both high and low frequency modeling, we introduce the sub-band reconstruction technique. Experimental results demonstrate that our \tool achieves SOTA performance of unified audio modeling compared to the single-codebook codes with only a 40 token rate.

\normalem
\bibliographystyle{IEEEbib}
\bibliography{references}

\end{document}